\documentclass[aps,prl,twocolumn,superscriptaddress]{revtex4-1}

\usepackage{colortbl}
\usepackage{amsmath}
\usepackage{amssymb}
\usepackage{gensymb}

\usepackage{natbib}

\usepackage{bm}

	\usepackage{amsmath}
	\usepackage{makeidx}
	\usepackage{amsfonts}
	\usepackage[ansinew]{inputenc}
	\usepackage[usenames,dvipsnames]{pstricks}
	\usepackage{subfigure}
	\usepackage{epsfig}
	\usepackage{pst-grad} 
	\usepackage{pst-plot} 
	\usepackage[colorlinks,hyperindex]{hyperref}
	\usepackage{tikz}
	\hypersetup
	{
		colorlinks,%
		citecolor=black,%
		linkcolor=black,%
		urlcolor=black,%
	}

\usepackage{gensymb}

\newcommand{\beq}{\begin{equation}}
\newcommand{\eeq}{\end{equation}}

\newcommand{\bb}[1]{\textcolor{black}{#1}}

\makeindex

\begin{document}
\title{Kilotesla plasmoid formation by a trapped relativistic laser beam}

\author{M.~Ehret}
\email[]{michael.ehret@u-bordeaux.fr}
\affiliation{Universit\'{e} de Bordeaux, CNRS, CEA, CELIA (Centre Lasers Intenses et Applications), UMR 5107, Talence, France}
\affiliation{Institut f\"{u}r Kernphysik, Technische Universit\"{a}t Darmstadt, Darmstadt, Germany}

\author{Yu.~ Kochetkov} 
\affiliation{National Research Nuclear University MEPhI, Moscow, Russian Federation}

\author{ Y.~Abe}
\affiliation{Institute of Laser Engineering, Osaka University, Japan}
\author{ K.~F.~F.~Law}
\affiliation{Institute of Laser Engineering, Osaka University, Japan}

\author{V.~ Stepanischev} 
\affiliation{National Research Nuclear University MEPhI, Moscow, Russian Federation}

\author{ S.~Fujioka}
\affiliation{Institute of Laser Engineering, Osaka University, Japan}
\author{E.~d'Humi\`eres}
\affiliation{Universit\'{e} de Bordeaux, CNRS, CEA, CELIA (Centre Lasers Intenses et Applications), UMR 5107, Talence, France}

\author{B.~Zielbauer}
\affiliation{Plasma Physik/PHELIX, GSI Helmholtzzentrum f\"{u}r Schwerionenforschung GmbH, Darmstadt,
Germany}

\author{V.~Bagnoud}
\affiliation{Plasma Physik/PHELIX, GSI Helmholtzzentrum f\"{u}r Schwerionenforschung GmbH, Darmstadt,
Germany}

\author{G.~Schaumann}
\affiliation{Institut f\"{u}r Kernphysik, Technische Universit\"{a}t Darmstadt, Darmstadt, Germany}

\author{M.~Roth}
\affiliation{Institut f\"{u}r Kernphysik, Technische Universit\"{a}t Darmstadt, Darmstadt, Germany}

\author{V. Tikhonchuk}
\affiliation{Universit\'{e} de Bordeaux, CNRS, CEA, CELIA (Centre Lasers Intenses et Applications), UMR 5107, Talence, France}
\affiliation{ELI-Beamlines, Institute of Physics Academy of Sciences of the Czech Republic, Doln\' i B\v re\v zany, Czech Republic}

\author{J.J.~Santos}
\email{joao.santos@u-bordeaux.fr }
\affiliation{Universit\'{e} de Bordeaux, CNRS, CEA, CELIA (Centre Lasers Intenses et Applications), UMR 5107, Talence, France}

\author{Ph.~ Korneev} 
\email{ph.korneev@gmail.com}
\affiliation{National Research Nuclear University MEPhI, Moscow, Russian Federation}
\affiliation{P.~N.~Lebedev Physical Institute of RAS, Moscow, Russian Federation}

\date{\today}

\begin{abstract}
A strong quasi-stationary magnetic field is generated in hollow targets with curved internal surface under the action of a relativistically intense picosecond laser pulse. Experimental data evidence formation of quasistationary strongly magnetized plasma structures decaying on the hundred picoseconds time scale, with the maximum value of magnetic field strength of the kilotesla scale. Numerical simulations unravel the importance of transient processes during the magnetic field generation, and suggest the existence of fast and slow regimes of plasmoid evolution depending on the interaction parameters. The principal setup is universal for perspective highly magnetized plasma application and fundamental studies.
\end{abstract}


\maketitle




Since the invention of chirp laser pulse amplification \cite{Strickland-oc1985}, optical methods became the most convenient tool in high energy density plasma studies. Optically driven magnetic fields are widely used for collimation of high energy particles \cite{Bailly-Grandvaux2016, Kar-nc2016, Fujioka-pp16}, inertial confinement fusion \cite{chang-prl11, Davies2017}, laboratory studies of astrophysically-relevant phenomena, including reconnection \cite{Yamada1999, Bulanov2017, Law-arxiv2019} and collisionless shock generation \cite{Schaeffer2017c, li-prl19}. In many cases optical schemes are preferrable than the ones based on external high-voltage sources fed by capacitor banks, where the magnetic field strength is limited to few tens Tesla in cm$^3$ scale volume on a $\mu$s time scale \cite{Higginson-hedp2015}. Optically-driven schemes normally use discharges supported by strong currents of accelerated electrons, induced by laser radiation. For example, in the capacitor-coil targets \cite{Korobkin-stpl79, Courtois-jap05, Santos-pop2018}, the laser pulse induces a high potential in the capacitor, which drives a strong discharge current through the coil. The nanosecond time scale for the discharge is set by the laser pulse duration and the target impedance. In similar experiments with ps laser pulses, the discharge current propagates across the target as a pulse, resulting in fast transient phenomena \cite{Kar-nc2016, Eh:2017}. The temporal evolution of the current is characterized by the aspect ratio $\varkappa=c \tau/L$, where $c$ is the light velocity, $\tau$ is the laser pulse duration, and $L$ is the characteristic length of the discharge coil. For ns laser pulses and mm-size coils the parameter $\varkappa$ is large, that corresponds to the ``slow'' regime of the magnetic field generation. For shorter pulses, in order to obtain a homogeneous and quasistatic magnetic field, the target size needs to be reduced. By miniaturization of targets it is possible to maintain the condition $\varkappa\gg1$ even for few-tens fs laser pulses \cite{Brantov-lpl2019}. In the scheme detailed below, based on the  micro-cavity ``snail'' targets, the parameter $\varkappa\sim0.5$ corresponds to an oscillatory regime of the laser-induced discharge. However, owing to the target geometry and a relatively high value of the parameter $\varkappa$, after a short transient regime, a strong quasistationary magnetic field was observed in the ablated plasma, reaching the kilotesla scale and then slowly decaying in time.
\begin{figure}
\centering\includegraphics[width=0.8\linewidth]{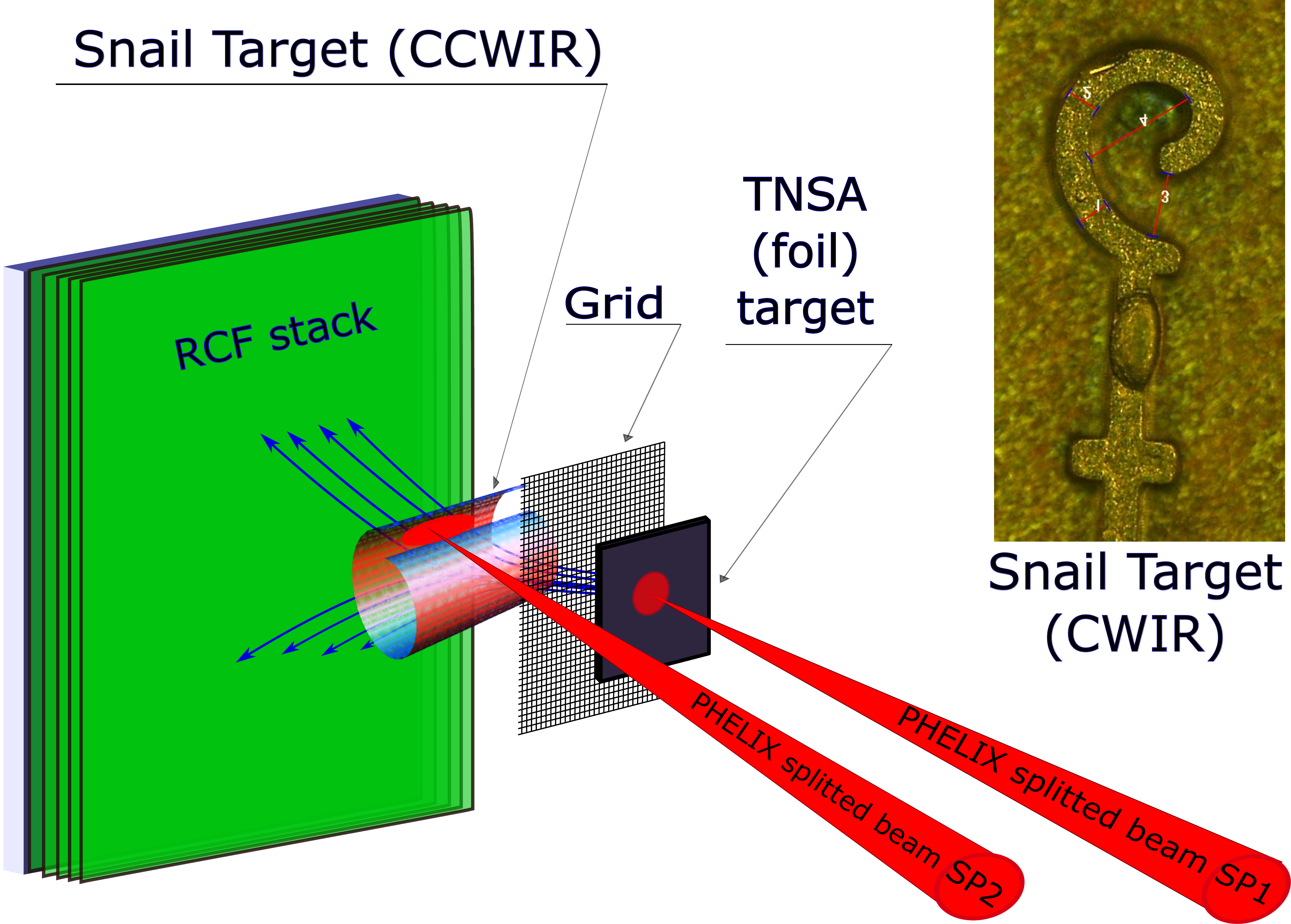}
\caption{Sketch of the experimental setup for the magnetic field generation generation (beam SP2) and protom backlighting (beam SP1). The photographic image of the CCWIR ``snail'' target is shown aside.}
\label{phelix_1}
\end{figure}
Those targets, proposed in Ref. \cite{Korneev-pre15} demonstrate for $\varkappa\sim 5$ a quasistationary behavior with a multi-kT scale magnetic field with an energetic efficiency reaching several percents. The theoretical modelling confirms the high robustness of these targets for magnetic field generation \cite{Korneev-jophcs2017}. They are driven by the discharge current, which normally dominates, and the current of relativistically accelerated electrons along the target surface \cite{Korneev-arxiv17, Abe2018a}. Depending of the geometrical shape, the magnetized structure in the cavity may be monopolar or bipolar, which alows to produce and study highly relativistic reconnection events \cite{Law-arxiv2019}.  
\begin{figure}[h]
\centering\includegraphics[width=\linewidth]{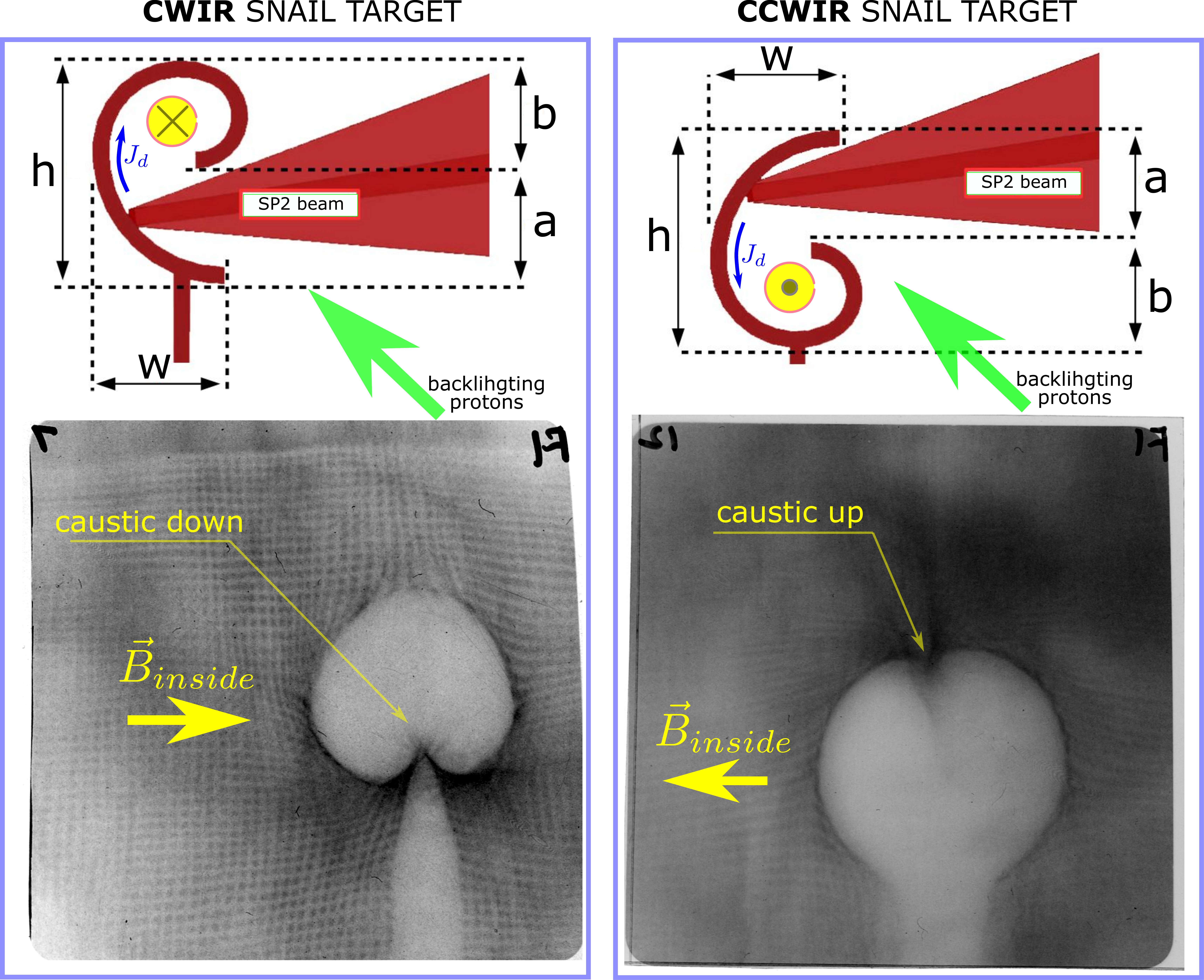}
\caption{Top: scheme of the Clock-Wise with Inlet Right (CWIR, left panel) and Counter-Clock-Wise with Inlet Right (CCWIR, right panel) targets, irradiated by the SP2 beam. Bottom: proton deflectometry images with the opposite position of the characteristic caustic features, the image is taken 64 ps for the CWIR and 80.5 ps for the CCWIR targets after the laser pulse. The experimentally defined direction of magnetic field lines is shown with yellow arrows.} 
\label{CCWIR+CWIR}
\end{figure}

Compared to the theoretical model \cite{Korneev-pre15}, in the present experiment, the target was bigger, while the laser pulse was shorter. The experiment was performed at the PHELIX laser facility at GSI (Darmstadt, Germany). The laser pulse at wavelength of $1056$ nm and duration of $0.5$ ps, was divided into two beams SP1 and SP2, each carrying energy of $50$ J. {The beams were tightly focused to focal spot of $\approx10~\mu$m FWHM (Full Width at Half Maximum) providing intensity $\approx2\times 10^{19}$ W/cm$^2$ at the target surface with the contrast $\sim 10^{-10}$. The SP1 beam was focused on a thin (4 $\mu$m) gold foil, serving as a source for backlighting TNSA (Target Normal Sheath Accelerated) energetic protons from the contamination layer \cite{Mackinnon2001}. The SP2 beam was focused on the internal surface of the ``snail'' target. Because of the grazing incidence, the intensity on target was $\lesssim10^{19}$ W/cm$^2$}. The experimental setup is illustrated by  Fig. \ref{phelix_1}.   
\bb{Targets were produced in the Technical University Darmstadt by using the MEMS (Micro-Electro-Mechanical Systems) technology.} Two types of targets were prepared for the experiment, the Clock-Wise with Inlet Right target (CWIR) and the Counter-Clock-Wise with Inlet Right target (CCWIR). The laser focusing scheme and the direction of the main discharge current $J_d$ are shown in Fig. \ref{CCWIR+CWIR}. These two different geometries were chosen so that the magnetic field, generated by the discharge currents, would have opposite polarities. The target image is shown in the inset panel of Fig. \ref{phelix_1}. The target geometry was designed and adjusted to fit the two main requirements, namely the laser pulse focusing cone geometry and the technological possibilities. Because of the tight focusing, the entrance for the laser beam had to be as wide as $a=136~\mu$m, half of the total target height $h=272~\mu$m. The width was $w=156~\mu$m in the plane of Fig. \ref{CCWIR+CWIR} and $100~\mu$m in the normal direction, i.e. along the target axis. 
 \begin{figure}[h]
\centering\includegraphics[width=\linewidth]{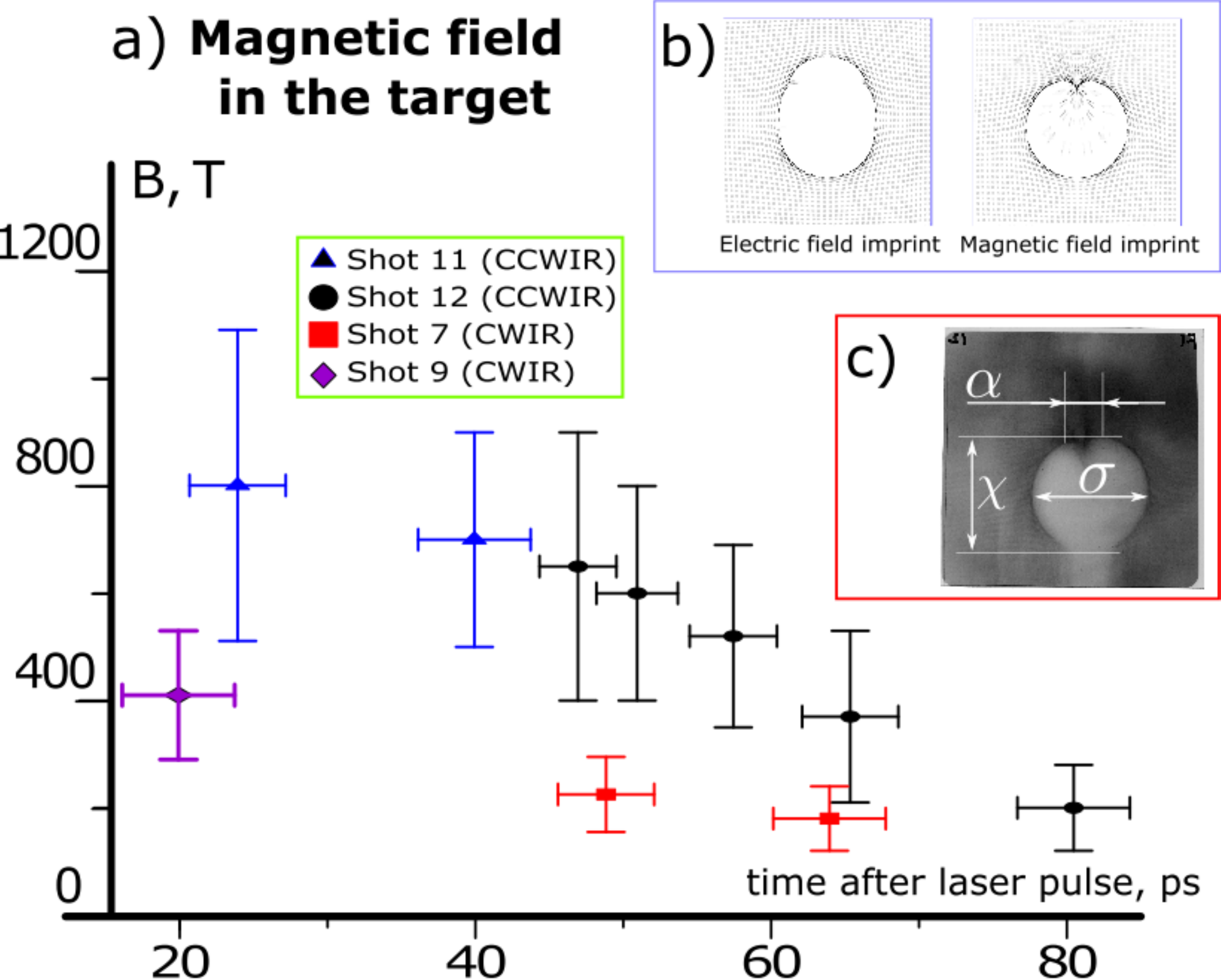}
\caption{a) Temporal evolution of the magnetic field in the target, estimated from the experimental data; b) synthetic proton radiagraphs for purely electric and magnetic fields in the target region; c) parameters $\alpha$, $\chi$, $\sigma$, used for adjustment of the test proton image on the RCF stack. 
 }
\label{B_t}
\end{figure}

The main result in this experiment was produced by the proton deflectometry diagnostics driven by SP1 beam. \bb{The maximum proton energy was around 10...15 MeV, which corresponds to the time of flight of $\sim50$ ps for the distance between the foil and the ``snail'' target of 2.8 mm. Additionally, SP1 was delayed for $\pm 500$ ps to study the time-dependence of the fields. Test protons were passing through the mesh, installed between the target and the foil. Multi-layer Radio-Chromic Film (RCF) stacks were used as a detector to visualize the fields spatial distribution and time evolution. Ballistic spreading of protons of different energies in flight provided the temporal resolution of the order of a few ps in successive RCF layers due to the Bragg peak energy absorption. The distance between the target and the RCF of 111 mm corresponds to the magnification factor of 40 in the RCF images.} Two examples of RCF images for the CWIR and CCWIR targets are shown in Fig. \ref{CCWIR+CWIR}. The characteristic caustic feature of the magnetic field solenoidal strucure is very similar to that observed in the capacitor-coil targets with a similar diagnostics \cite{Santos-pop2018}. However, in difference from Ref. \cite{Santos-pop2018}, in the present setup, the magnetic field is generated in a hot plasma, but not in a vacuum. 

Analysis of the proton deflectometry data was performed by using the test-particle approach.  The proton source is considered as a point source and the space-charge effects in the probe beam are neglected, so the deviations of proton trajectories are caused solely by the interaction with the fields generated in the target. It was verified, that the protons with trajectories passing through the coil, do not contribute to the RCF image. That test removes the question of a possible interaction of the test protons with the target material.  Reconstruction of fields from the RCF images is a complicated multi-dimensional inverse problem. Each image was characterized by three parameters $\alpha$, $\chi$ and $\sigma$ as shown in Fig \ref{B_t} c). Field maps were obtained with a resolution of 10 $\mu$m. The electric field was defined for a conducting snail target with a real geometry, charged to a parametrically adjustable potential. Magnetic field was calculated using the software Radia 4.3 \cite{Chubar1998}, \bb{for an electrical current in a coil-like circuit, assumed to be inside the target, close to its internal surface. The circuit was connected straight in the slit, and had parametrically adjustable inner and outer radii, as well as the straight section length and the total height.}
A proton trajectory through the field was calculated by solving the Newton's equation of motion with the Lorentz force $\vec F=e(\vec E+[\vec v\times \vec B])$ with a 0.1 ps time resolution.

First, the synthetic images were constructed assuming only electric or only magnetic field. Within the whole space of parameters, in the case of only electric field, the synthetic images have a smooth oval shape without the point-like caustics. In contrast, for the case of only magnetic field, there was always a point-like singularity, see Fig. \ref{B_t} b). Patterns similar to the experimental data, can be produced only assuming the presence of both the electric field and magnetic field of several hundreds of Tesla.
To determine the magnetic field value, the target's electric potential $U$ and current in the simulation were parametrically varied for possible current circuit shapes to create 3-dimensional plots: $\alpha(B, U)$, $\chi(B, U)$ and $\sigma(B, U)$, where $B$ is the magnetic field value in the center of the coil. These plots were reduced to three curves by using experimental data for the dependence  $U(B)$ . The intersections for the pairs of the curves should coincide. This condition appears to be quite sensitive to the curcuit shape and allows to make the initial estimate for the circuit parameters and the size of the plasmoid, minimizing the distance between the intersection points.
\begin{figure}[h]
	\centering\includegraphics[width=\linewidth]{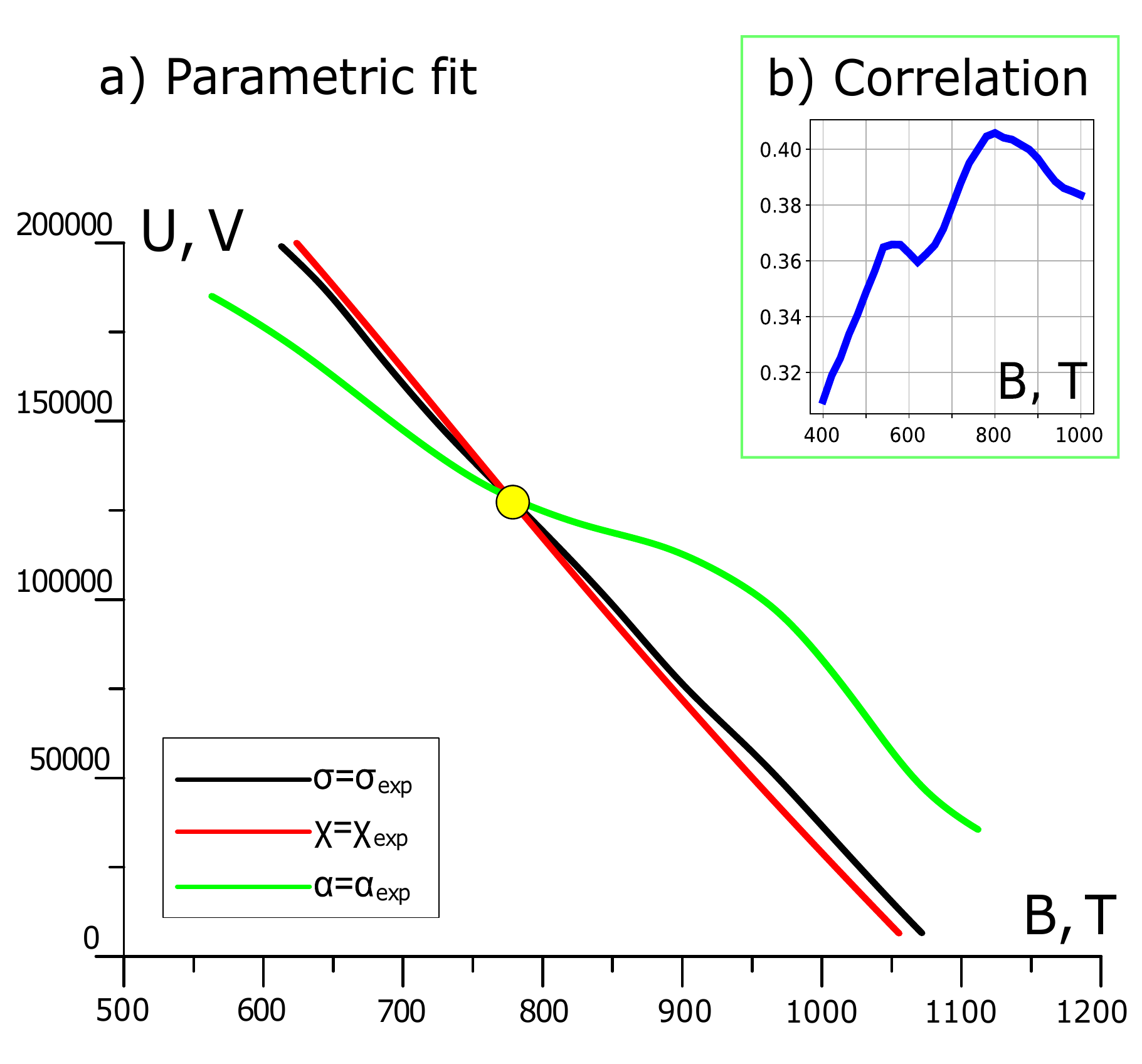}
	\caption{Panel a): illustration of the parametric fit to the experimental data for three curves $\sigma(U,B)$, $\chi(U,B)$ and $\alpha(U,B)$. Panel b): correlation coefficient between experimental and synthetic image dependence on magnetic field value in the target. Shot \#11 at 24 ps.}
	\label{Correlate}
\end{figure}
To define the parameters more precisely, an additional correlation analysis was carried out. The circuit parameters and current were varied around the initially estimated values. Afterwards, for each synthetic image a structural similarity index, the normalized cross-correlation matrix and thus the correlation coefficient were calculated to determine the best fitting image and consequently the corresponding circuit parameters and the value of the magnetic field (see Fig. \ref{Correlate}).

With the method described above, the magnetic field was defined for different energies of the backlight protons, i.e. for different time moments. The results are shown in Fig. \ref{B_t}). The maximum value of the magnetic field observed is (800$\pm$280) T at 24 ps after the laser pulse. Then, the magnetic field decays to $\approx200\pm$80 T at $\approx80$ ps. This time corresponds to the target expansion, leading to the loss of confinement. During its temporal evolution, the estimated radius of the current circuit, corresponding to the plasmoid size, was found to change from 70 to 100 $\mu$m, and height - from 70 $\mu$m to 120 $\mu$m. The energy conversion efficiency $\eta$ of the magnetic field generation, was estimated as a relation of the magnetic field energy $W_B$ to the  of the total laser energy, delivered to the target $W_L\approx 50$ J. At maximum it was $\eta\approx$ 4\% ($W_B\sim 2$ J). This value is comparable, although somewhat less, to $\sim$10\%, estimated for the capacitor-coil targets \cite{Santos-NJP2015}, but target optimization may result in a higher efficiency. Note, that due to the compactization of the target the kilotesla scale of the magnetic field was obtained with only 50 J of the laser beam.
  
 \begin{figure}[h]
\centering\includegraphics[width=\linewidth]{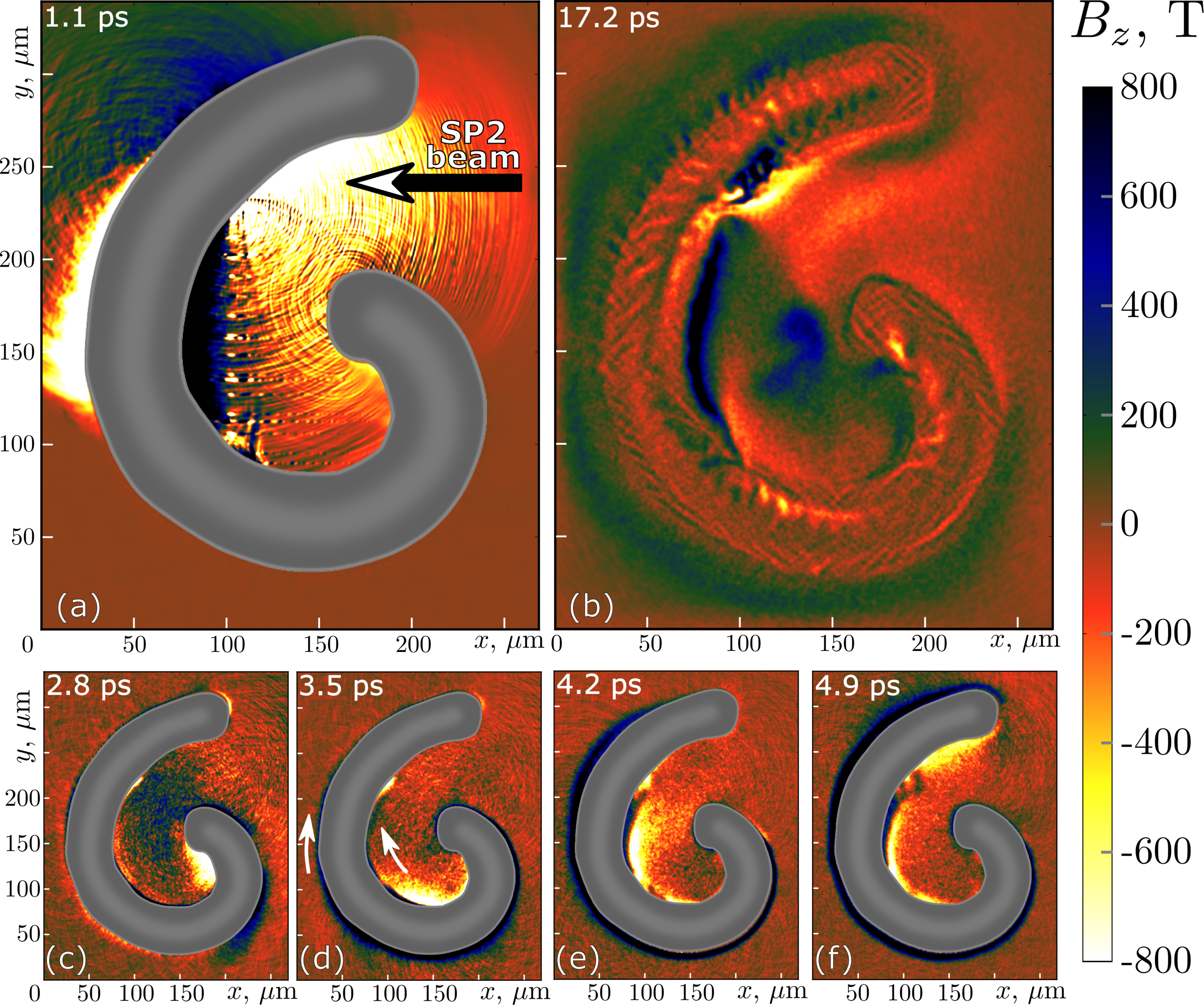}
\caption{2D simulation of interaction of the PHELIX SP2 laser beam with a real-size target: a) first stage of interaction at 1.1 ps, laser pulse just entered the target; b) at time of 17.2 ps inside the target a magnetized plasmoid is formed; c)-f) a transient discharge wave is travelling along the target, the white arrows in panel d) show direction of propagation of the wave. The target volume in panels a), c)-f) is filled with the grey color, corresponding to the initial plasma density, the light internal region inside the target bulk is the reduced 10 times the maximum density to save the calculation time.}
\label{BPIC}
	\end{figure}

The magnetic field generation was simulated in 2D with a Particle-in-Cell (PIC) code PICLS \cite{Sentoku-jcp08} with the real target size and geometry. Simulations were performed with a plasma made of electrons and singly charged ions with atomic mass A=64 corresponding to copper. The electron density was $10$ times the critical plasma density. Simulation box size was $4080\times5152$ cells or $269\times340~\mu$m, with 50 electrons and 50 ions per cell. The time step was 0.2 fs and the total simulation time was about 17 ps. Inside the bulk of the target where the field does not penetrate, to save computation time and suppress numerical heating, the density was reduced, see the grey target-shaped figures in panels a), c)-f) in Fig. \ref{BPIC}. The magnetic field amplitude $B_z$ is shown at time moments 1.1 ps (panel a), initial interaction stage, the SP2 beam just entered the target inlet), and 17.2 ps (panel b), the latest simulation time). The fields are averaged spatially on the scale of 0.5 $\mu$m with a Gaussian envelope to reduce the visual noise. 
The laser pulse length is less than the target perimeter ($\varkappa\approx0.5<1$). This results in a premature disruption of the discharge currents, which is clearly observed in the simulations. In panels c) - f) in Fig. \ref{BPIC} a transient process of fast propagation of the discharge wave is shown. The wave has a complicated shape, containing a positive and a negative parts, which originate from the laser focusing point. The two discharge pulses with opposite polarities are travelling along the target surface, meeting each other at some point on the other side and propagating further. After a transient time interval, a quasistationary plasmoid is formed inside the target cavity. The magnetic field value of the order of 600 - 800 T, is quite similar to the experimental values. The important difference between the simulation and the experiment is the holder, which may change the conditions of the discharge evolution. The effect of the holder on the final magnetic field value may be observed in the experimental data in Fig. \ref{B_t}. It is seen, that for the same time, in both CWIR shots (\#7 and \#9) the field value is substantially less, than in both CCWIR shots (\#11 and \#12). This probably is the result of stalk grounding, which in the CCWIR target increases electric capacitance.

This experiment demonstrates that small ``snail'' targets driven by high-intensity ps-short laser pulses, generate much stronger quasi-static magnetic fields than bigger targets with longer, sub-ns pulses \cite{Pisarczyk-scirep2018}, due to a higher energy density in this regime. Compared to the capacitor-coil targets  \cite{Santos-pop2018}, the snail targets produce stronger magnetic fields but in a smaller volume and for a shorter time. The most important difference consists in the fact that magnetic field, after a short transient period, is frozen in the collisionless plasma. That self-organized localized structure of a plasma blob and a magnetic field - plasmoid - exists at the hydrodynamic time scale of the order of hundred of ps. During this period, it presents a medium where strong coupling between an intense laser beam and magnetic field may be realized. Recent studies show strong effect of nonlinear plasma dynamics \cite{Bulanov-pop2013, rassou-pop15}, on the increase of charge and energy of accelerated electrons in laser wakefield acceleration \cite{Hur-phla2008, Drouin-jmp2012, Jha-prstab2012} and in other schemes \cite{Kant-hedp2016}. Extremely strong magnetization may considerably facilitate proton collimation emitted from near-critical \cite{Kuri-lpb2018} and overdense plasmas \cite{Kuri-pop2017}. Unprecendently high magnetization, obtained in a hot collisionless palsmoid, becide applications in particle acceleration and radiation generation, paves the way to fundamental studies of collisionless magnetized plasmas \cite{Law-arxiv2019}.      

In conclusion, a fast ps-scale formation of a highly magnetized plasma structure is observed. It is shown that, being driven by an optically-induced discharge, it is governed by a parameter $\varkappa$ relating a target size and laser beam length. For $\varkappa\lesssim 1$, the process demonstrates transient features, which relax to a stable on a hydrodynamic timescale strongly-magnetized plasmoid, evidenced experimentally.  Magnetic field strength reached the value of 800$\pm$280 T at maximum for only 50 J of the invested laser energy. The presented scheme allow the robust strongly magnetized plasma production for multiple applications in fundamental physics studies, including laboratory astrophysics experiments.

\begin{acknowledgments}
\textit{Acknowledgements}\\
The authors are grateful to the PHELIX team, for support of the experimental campaign. Numerical simulations were supported in part by the HPC resources of CINES under the allocations DARI \#A0040507594 and DARI \#A0050506129 made by GENCI (Grand Equipment National de Calcul Intensif) and in part by resources of NRNU MEPhI High-Performance Computing Center. The work was partially supported by the MEPhI Academic Excellence Project (contract No. 02.a03.21.0005, 27.08.2013) and by the ``Investments for the future'' program IdEx Bordeaux LAPHIA (No. ANR-10-IDEX-03-02). This research was also carried out within the framework of the EUROfusion Consortium. Y. A., K. F. F. L., and S. F. were supported by fundings from JSPS-MEXT by Grant-in-Aid for JSPS Research Fellow (18J11119, 18J11354), Grants-in-Aid, KAKENHI (Grant No. 15KK0163, 15K21767, 16K13918, 16H02245) and Bilateral Program for Supporting International Joint Research.
\end{acknowledgments}

\bibliography{library} 

\begin{thebibliography}{34}
\providecommand{\natexlab}[1]{#1}
\providecommand{\url}[1]{\texttt{#1}}
\expandafter\ifx\csname urlstyle\endcsname\relax
  \providecommand{\doi}[1]{doi: #1}\else
  \providecommand{\doi}{doi: \begingroup \urlstyle{rm}\Url}\fi

\bibitem[Strickland and Mourou(1985)]{Strickland-oc1985}
Donna Strickland and Gerard Mourou.
\newblock {Compression of amplified chirped optical pulses}.
\newblock \emph{Optics Communications}, 55\penalty0 (6):\penalty0 447--449, oct
  1985.
\newblock ISSN 00304018.
\newblock \doi{10.1016/0030-4018(85)90151-8}.
\newblock URL
  \url{http://linkinghub.elsevier.com/retrieve/pii/0030401885901518}.

\bibitem[Bailly-Grandvaux et~al.(2018)Bailly-Grandvaux, Santos, Bellei,
  Forestier-Colleoni, Fujioka, Giuffrida, Honrubia, Batani, Bouillaud, Chevrot,
  Cross, Crowston, Dorard, Dubois, Ehret, Gregori, Hulin, Kojima, Loyez,
  Marqu{\`{e}}s, Morace, Nicola{\"{i}}, Roth, Sakata, Schaumann, Serres,
  Servel, Tikhonchuk, Woolsey, and Zhang]{Bailly-Grandvaux2016}
M.~Bailly-Grandvaux, J.~J. Santos, C.~Bellei, P.~Forestier-Colleoni,
  S.~Fujioka, L.~Giuffrida, J.~J. Honrubia, D.~Batani, R.~Bouillaud,
  M.~Chevrot, J.~E. Cross, R.~Crowston, S.~Dorard, J.-L. Dubois, M.~Ehret,
  G.~Gregori, S.~Hulin, S.~Kojima, E.~Loyez, J.-R. Marqu{\`{e}}s, A.~Morace,
  Ph. Nicola{\"{i}}, M.~Roth, S.~Sakata, G.~Schaumann, F.~Serres, J.~Servel,
  V.~T. Tikhonchuk, N.~Woolsey, and Z.~Zhang.
\newblock {Guiding of relativistic electron beams in dense matter by
  laser-driven magnetostatic fields}.
\newblock \emph{Nature Communications}, 9\penalty0 (1):\penalty0 102, dec 2018.
\newblock ISSN 2041-1723.
\newblock \doi{10.1038/s41467-017-02641-7}.
\newblock URL \url{http://arxiv.org/abs/1608.08101
  http://www.nature.com/articles/s41467-017-02641-7}.

\bibitem[Kar et~al.(2016)Kar, Ahmed, Prasad, Cerchez, Brauckmann, Aurand,
  Cantono, Hadjisolomou, Lewis, Macchi, Nersisyan, Robinson, Schroer,
  Swantusch, Zepf, Willi, and Borghesi]{Kar-nc2016}
Satyabrata Kar, Hamad Ahmed, Rajendra Prasad, Mirela Cerchez, Stephanie
  Brauckmann, Bastian Aurand, Giada Cantono, Prokopis Hadjisolomou, Ciaran L~S
  Lewis, Andrea Macchi, Gagik Nersisyan, Alexander P~L Robinson, Anna~M.
  Schroer, Marco Swantusch, Matt Zepf, Oswald Willi, and Marco Borghesi.
\newblock {Guided post-acceleration of laser-driven ions by a miniature modular
  structure}.
\newblock \emph{Nature Communications}, 7:\penalty0 10792, apr 2016.
\newblock ISSN 20411723.
\newblock \doi{10.1038/ncomms10792}.
\newblock URL \url{http://www.nature.com/doifinder/10.1038/ncomms10792}.

\bibitem[Fujioka et~al.(2016)Fujioka, Arikawa, Kojima, Johzaki, Nagatomo,
  Sawada, Lee, Shiroto, Ohnishi, Morace, Vaisseau, Sakata, Abe, Matsuo, {Farley
  Law}, Tosaki, Yogo, Shigemori, Hironaka, Zhang, Sunahara, Ozaki, Sakagami,
  Mima, Fujimoto, Yamanoi, Norimatsu, Tokita, Nakata, Kawanaka, Jitsuno,
  Miyanaga, Nakai, Nishimura, Shiraga, Kondo, Bailly-Grandvaux, Bellei, Santos,
  and Azechi]{Fujioka-pp16}
Shinsuke Fujioka, Yasunobu Arikawa, Sadaoki Kojima, Tomoyuki Johzaki, Hideo
  Nagatomo, Hiroshi Sawada, Seung~Ho Lee, Takashi Shiroto, Naofumi Ohnishi,
  Alessio Morace, Xavier Vaisseau, Shohei Sakata, Yuki Abe, Kazuki Matsuo,
  King~Fai {Farley Law}, Shota Tosaki, Akifumi Yogo, Keisuke Shigemori,
  Yoichiro Hironaka, Zhe Zhang, Atsushi Sunahara, Tetsuo Ozaki, Hitoshi
  Sakagami, Kunioki Mima, Yasushi Fujimoto, Kohei Yamanoi, Takayoshi Norimatsu,
  Shigeki Tokita, Yoshiki Nakata, Junji Kawanaka, Takahisa Jitsuno, Noriaki
  Miyanaga, Mitsuo Nakai, Hiroaki Nishimura, Hiroyuki Shiraga, Kotaro Kondo,
  Mathieu Bailly-Grandvaux, Claudio Bellei, Jo{\~{a}}o~Jorge Santos, and
  Hiroshi Azechi.
\newblock {Fast ignition realization experiment with high-contrast kilo-joule
  peta-watt LFEX laser and strong external magnetic field}.
\newblock \emph{Phys. Plasmas}, 23\penalty0 (5):\penalty0 56308, may 2016.
\newblock ISSN 10897674.
\newblock \doi{10.1063/1.4948278}.
\newblock URL
  \url{http://scitation.aip.org/content/aip/journal/pop/23/5/10.1063/1.4948278
  http://aip.scitation.org/doi/10.1063/1.4948278
  http://dx.doi.org/10.1063/1.4948278 http://aip.scitation.org/toc/php/23/5}.

\bibitem[Chang et~al.(2011)Chang, Fiksel, Hohenberger, Knauer, Betti, Marshall,
  Meyerhofer, S{\'{e}}guin, and Petrasso]{chang-prl11}
P~Y Chang, G~Fiksel, M~Hohenberger, J~P Knauer, R~Betti, F~J Marshall, D~D
  Meyerhofer, F~H S{\'{e}}guin, and R~D Petrasso.
\newblock {Fusion Yield Enhancement in Magnetized Laser-Driven Implosions}.
\newblock \emph{Physical Review Letters}, 107\penalty0 (3):\penalty0 035006,
  jul 2011.
\newblock ISSN 0031-9007.
\newblock \doi{10.1103/PhysRevLett.107.035006}.
\newblock URL \url{http://link.aps.org/doi/10.1103/PhysRevLett.107.035006
  https://link.aps.org/doi/10.1103/PhysRevLett.107.035006}.

\bibitem[Davies et~al.(2017)Davies, Barnak, Betti, Campbell, Chang, Sefkow,
  Peterson, Sinars, and Weis]{Davies2017}
J.~R. Davies, D.~H. Barnak, R.~Betti, E.~M. Campbell, P.~Y. Chang, A.~B.
  Sefkow, K.~J. Peterson, D.~B. Sinars, and M.~R. Weis.
\newblock {Laser-driven magnetized liner inertial fusion}.
\newblock \emph{Physics of Plasmas}, 24\penalty0 (6):\penalty0 1--6, 2017.
\newblock ISSN 10897674.
\newblock \doi{10.1063/1.4984779}.

\bibitem[Yamada(1999)]{Yamada1999}
Masaaki Yamada.
\newblock {Review of controlled laboratory experiments on physics of magnetic
  reconnection}.
\newblock \emph{Journal of Geophysical Research}, 104\penalty0 (A7):\penalty0
  14529, 1999.
\newblock ISSN 0148-0227.
\newblock \doi{10.1029/1998JA900169}.

\bibitem[Bulanov(2017)]{Bulanov2017}
S.~V. Bulanov.
\newblock {Magnetic reconnection: from MHD to QED}.
\newblock \emph{Plasma Physics and Controlled Fusion}, 59\penalty0
  (1):\penalty0 014029, 2017.
\newblock ISSN 0741-3335.
\newblock \doi{10.1088/0741-3335/59/1/014029}.
\newblock URL
  \url{http://stacks.iop.org/0741-3335/59/i=1/a=014029?key=crossref.e23730d32d20fe6dd737380d0dbe765b
  http://arxiv.org/abs/1610.01873}.

\bibitem[Law et~al.(2019)Law, Abe, Morace, Arikawa, Sakata, Lee, Matsuo,
  Morita, Ochiai, Liu, Yogo, Okamoto, Golovin, Ehret, Ozaki, Nakai, Sentoku,
  Santos, D'Humi{\`{e}}res, Korneev, and Fujioka]{Law-arxiv2019}
K.~F.~F. Law, Y.~Abe, A.~Morace, Y.~Arikawa, S.~Sakata, S.~Lee, K.~Matsuo,
  H.~Morita, Y.~Ochiai, C.~Liu, A.~Yogo, K.~Okamoto, D.~Golovin, M.~Ehret,
  T.~Ozaki, M.~Nakai, Y.~Sentoku, J.~J. Santos, E.~D'Humi{\`{e}}res, Ph.
  Korneev, and S.~Fujioka.
\newblock {Hard particle spectra of galactic X-ray sources by relativistic
  magnetic reconnection in laser lab}.
\newblock \emph{arXiv}, apr 2019.
\newblock URL \url{https://arxiv.org/abs/1904.02850
  http://arxiv.org/abs/1904.02850}.

\bibitem[Schaeffer et~al.(2017)Schaeffer, Fox, Haberberger, Fiksel,
  Bhattacharjee, Barnak, Hu, Germaschewski, and Follett]{Schaeffer2017c}
D.~B. Schaeffer, W.~Fox, D.~Haberberger, G.~Fiksel, A.~Bhattacharjee, D.~H.
  Barnak, S.~X. Hu, K.~Germaschewski, and R.~K. Follett.
\newblock {High-Mach number, laser-driven magnetized collisionless shocks}.
\newblock \emph{Physics of Plasmas}, 24\penalty0 (12), 2017.
\newblock ISSN 10897674.
\newblock \doi{10.1063/1.4989562}.

\bibitem[Li et~al.(2019)Li, Tikhonchuk, Moreno, Sio, D'Humi{\`{e}}res, Ribeyre,
  Korneev, Atzeni, Betti, Birkel, Campbell, Follett, Frenje, Hu, Koenig,
  Sakawa, Sangster, Seguin, Takabe, Zhang, and Petrasso]{li-prl19}
C.~K. Li, V.~T. Tikhonchuk, Q~Moreno, H~Sio, E.~D'Humi{\`{e}}res, X~Ribeyre,
  Ph~Korneev, S~Atzeni, R.~Betti, A.~Birkel, E.~M. Campbell, R.~K. Follett,
  J.~A. Frenje, S.~X. Hu, M.~Koenig, Y.~Sakawa, T.~C. Sangster, F.~H. Seguin,
  H~Takabe, S~Zhang, and R.~D. Petrasso.
\newblock {Collisionless Shocks Driven by Supersonic Plasma Flows with
  Self-Generated Magnetic Fields}.
\newblock \emph{Physical Review Letters}, 123\penalty0 (5):\penalty0 055002,
  jul 2019.
\newblock ISSN 0031-9007.
\newblock \doi{10.1103/PhysRevLett.123.055002}.
\newblock URL \url{https://link.aps.org/doi/10.1103/PhysRevLett.123.055002}.

\bibitem[Higginson et~al.(2015)Higginson, Korneev, B{\'{e}}ard, Chen,
  D'Humi{\`{e}}res, P{\'{e}}pin, Pikuz, Pollock, Riquier, Tikhonchuk, and
  Fuchs]{Higginson-hedp2015}
D.~P. Higginson, Ph~Korneev, J.~B{\'{e}}ard, S.~N. Chen, E.~D'Humi{\`{e}}res,
  H.~P{\'{e}}pin, S.~Pikuz, B.~Pollock, R.~Riquier, V.~Tikhonchuk, and
  J.~Fuchs.
\newblock {A novel platform to study magnetized high-velocity collisionless
  shocks}.
\newblock \emph{High Energy Density Physics}, 17:\penalty0 190--197, dec 2015.
\newblock ISSN 15741818.
\newblock \doi{10.1016/j.hedp.2014.11.007}.
\newblock URL
  \url{http://linkinghub.elsevier.com/retrieve/pii/S1574181814000858}.

\bibitem[Korobkin and Motylev(1979)]{Korobkin-stpl79}
V.~V. Korobkin and S.~L. Motylev.
\newblock {Laser method for producing strong magnetic fields}.
\newblock \emph{Soviet Technical Physics Letters (in Russian)}, 5:\penalty0
  474--476, 1979.

\bibitem[Courtois et~al.(2005)Courtois, Ash, Chambers, Grundy, and
  Woolsey]{Courtois-jap05}
C.~Courtois, A.~D. Ash, D.~M. Chambers, R.~A~D Grundy, and N.~C. Woolsey.
\newblock {Creation of a uniform high magnetic-field strength environment for
  laser-driven experiments}.
\newblock \emph{Journal of Applied Physics}, 98\penalty0 (5):\penalty0 54913,
  sep 2005.
\newblock ISSN 00218979.
\newblock \doi{10.1063/1.2035896}.
\newblock URL
  \url{http://link.aip.org/link/JAPIAU/v98/i5/p054913/s1{\%}7B{\&}{\%}7DAgg=doi
  http://scitation.aip.org/content/aip/journal/jap/98/5/10.1063/1.2035896
  http://aip.scitation.org/doi/10.1063/1.2035896}.

\bibitem[Santos et~al.(2018)Santos, Bailly-Grandvaux, Ehret, Arefiev, Batani,
  Beg, Calisti, Ferri, Florido, Forestier-Colleoni, Fujioka, Gigosos,
  Giuffrida, Gremillet, Honrubia, Kojima, Korneev, Law, Marqu{\`{e}}s, Morace,
  Moss{\'{e}}, Peyrusse, Rose, Roth, Sakata, Schaumann, Suzuki-Vidal,
  Tikhonchuk, Toncian, Woolsey, and Zhang]{Santos-pop2018}
J.~J. Santos, M.~Bailly-Grandvaux, M.~Ehret, A.~V. Arefiev, D.~Batani, F.~N.
  Beg, A.~Calisti, S.~Ferri, R.~Florido, P.~Forestier-Colleoni, S.~Fujioka,
  M.~A. Gigosos, L.~Giuffrida, L.~Gremillet, J.~J. Honrubia, S.~Kojima, Ph.
  Korneev, K.~F.~F. Law, J.-R. Marqu{\`{e}}s, A.~Morace, C.~Moss{\'{e}},
  O.~Peyrusse, S.~Rose, M.~Roth, S.~Sakata, G.~Schaumann, F.~Suzuki-Vidal,
  V.~T. Tikhonchuk, T.~Toncian, N.~Woolsey, and Z.~Zhang.
\newblock {Laser-driven strong magnetostatic fields with applications to
  charged beam transport and magnetized high energy-density physics}.
\newblock \emph{Physics of Plasmas}, 25\penalty0 (5):\penalty0 056705, may
  2018.
\newblock ISSN 1070-664X.
\newblock \doi{10.1063/1.5018735}.
\newblock URL \url{http://arxiv.org/abs/1712.07175
  http://aip.scitation.org/doi/10.1063/1.5018735}.

\bibitem[{M. Ehret, J.I. Apinaniz, M. Bailly-Grandvaux, V. Bagnoud, C. Brabetz,
  S. Malko} and {Morace, M. Roth, G. Schaumann, L. Volpe}(2017)]{Eh:2017}
A.~{M. Ehret, J.I. Apinaniz, M. Bailly-Grandvaux, V. Bagnoud, C. Brabetz, S.
  Malko} and J.J.~Santos {Morace, M. Roth, G. Schaumann, L. Volpe}.
\newblock {Energy selective focusing of TNSA beams by picosecond-laser driven
  ultra-fast EM fields}.
\newblock \emph{News and Reports from HEDgeHOB}, GSI-2017-2:\penalty0 19--20,
  2017.

\bibitem[Brantov et~al.(2019)Brantov, Korneev, and Bychenkov]{Brantov-lpl2019}
A.~V. Brantov, Ph. Korneev, and V.~Yu. Bychenkov.
\newblock {Magnetic field generation from a coil-shaped foil by a
  laser-triggered hot-electron current}.
\newblock \emph{Laser Physics Letters}, 16\penalty0 (6):\penalty0 066006, jun
  2019.
\newblock ISSN 1612-2011.
\newblock \doi{10.1088/1612-202X/ab1cb4}.
\newblock URL \url{http://arxiv.org/abs/1904.08673
  http://stacks.iop.org/1612-202X/16/i=6/a=066006?key=crossref.5b37fe35581dd7027c3d5a9d53233465}.

\bibitem[Korneev et~al.(2015)Korneev, D'Humi{\`{e}}res, and
  Tikhonchuk]{Korneev-pre15}
Ph. Korneev, E.~D'Humi{\`{e}}res, and V.~Tikhonchuk.
\newblock {Gigagauss-scale quasistatic magnetic field generation in a
  snail-shaped target}.
\newblock \emph{Physical Review E}, 91\penalty0 (4):\penalty0 43107, apr 2015.
\newblock ISSN 1539-3755.
\newblock \doi{10.1103/PhysRevE.91.043107}.
\newblock URL \url{https://journals.aps.org/pre/pdf/10.1103/PhysRevE.91.043107
  http://arxiv.org/abs/1410.0053 http://dx.doi.org/10.1103/PhysRevE.91.043107
  https://link.aps.org/doi/10.1103/PhysRevE.91.043107}.

\bibitem[Korneev(2017)]{Korneev-jophcs2017}
Philipp Korneev.
\newblock {Magnetized plasma structures in laser-irradiated curved targets}.
\newblock \emph{Journal of Physics: Conference Series}, 788\penalty0
  (1):\penalty0 012042, jan 2017.
\newblock ISSN 1742-6588.
\newblock \doi{10.1088/1742-6596/788/1/012042}.
\newblock URL
  \url{http://stacks.iop.org/1742-6596/788/i=1/a=012042?key=crossref.7338ac8a0ea6cbe7f5be21e3af5ef162}.

\bibitem[Korneev et~al.(2017)Korneev, Abe, Law, Bochkarev, Fujioka, Kojima,
  Lee, Sakata, Matsuo, Oshima, Morace, Arikawa, Yogo, Nakai, Norimatsu,
  D'Humi{\'{e}}res, Santos, Kondo, Sunahara, Bychenkov, Gus'kov, and
  Tikhonchuk]{Korneev-arxiv17}
Ph~Korneev, Y~Abe, K.~F.~F. Law, S~G Bochkarev, S~Fujioka, S~Kojima, S.~H. Lee,
  S~Sakata, K~Matsuo, A~Oshima, A~Morace, Y~Arikawa, A~Yogo, M~Nakai,
  T~Norimatsu, E.~D'Humi{\'{e}}res, J~J Santos, K~Kondo, A~Sunahara, V.~Yu.
  Bychenkov, S.~Gus'kov, and V~Tikhonchuk.
\newblock {Laser electron acceleration on curved surfaces}.
\newblock \emph{arXiv}, page 1711.00971, nov 2017.
\newblock URL \url{https://arxiv.org/pdf/1711.00971.pdf
  http://arxiv.org/abs/1711.00971}.

\bibitem[Abe et~al.(2018)Abe, Law, Korneev, Fujioka, Kojima, Lee, Sakata,
  Matsuo, Oshima, Morace, Arikawa, Yogo, Nakai, Norimatsu, D'Humi{\`{e}}res,
  Santos, Kondo, Sunahara, Gus'kov, and Tikhonchuk]{Abe2018a}
Y.~Abe, K.~F.~F. Law, Ph. Korneev, S.~Fujioka, S.~Kojima, S.-H. Lee, S.~Sakata,
  K.~Matsuo, A.~Oshima, A.~Morace, Y.~Arikawa, A.~Yogo, M.~Nakai, T.~Norimatsu,
  E.~D'Humi{\`{e}}res, J.~J. Santos, K.~Kondo, A.~Sunahara, S.~Gus'kov, and
  V.~Tikhonchuk.
\newblock {Whispering Gallery Effect in Relativistic Optics}.
\newblock \emph{JETP Letters}, 107\penalty0 (6):\penalty0 351--354, mar 2018.
\newblock ISSN 0021-3640.
\newblock \doi{10.1134/S0021364018060012}.
\newblock URL \url{http://link.springer.com/10.1134/S0021364018060012}.

\bibitem[Mackinnon et~al.(2001)Mackinnon, Borghesi, Hatchett, Key, Patel,
  Campbell, Schiavi, Snavely, Wilks, and Willi]{Mackinnon2001}
A.~J Mackinnon, M~Borghesi, S~Hatchett, M~H Key, P~K Patel, H~Campbell,
  A.~Schiavi, R~Snavely, S~C Wilks, and O~Willi.
\newblock {Effect of Plasma Scale Length on Multi-MeV Proton Production by
  Intense Laser Pulses}.
\newblock \emph{Physical Review Letters}, 86\penalty0 (9):\penalty0 1769--1772,
  feb 2001.
\newblock ISSN 0031-9007.
\newblock \doi{10.1103/PhysRevLett.86.1769}.
\newblock URL \url{https://link.aps.org/doi/10.1103/PhysRevLett.86.1769}.

\bibitem[Chubar et~al.(1998)Chubar, Elleaume, and Chavanne]{Chubar1998}
Oleg Chubar, Pascal Elleaume, and Joel Chavanne.
\newblock {A three-dimensional magnetostatics computer code for insertion
  devices}.
\newblock \emph{Journal of Synchrotron Radiation}, 5\penalty0 (3):\penalty0
  481--484, may 1998.
\newblock ISSN 0909-0495.
\newblock \doi{10.1107/S0909049597013502}.
\newblock URL \url{http://scripts.iucr.org/cgi-bin/paper?S0909049597013502}.

\bibitem[Santos et~al.(2015)Santos, Bailly-Grandvaux, Giuffrida,
  Forestier-Colleoni, Fujioka, Zhang, Korneev, Bouillaud, Dorard, Batani,
  Chevrot, Cross, Crowston, Dubois, Gazave, Gregori, D'Humi{\`{e}}res, Hulin,
  Ishihara, Kojima, Loyez, Marqu{\`{e}}s, Morace, Nicola{\"{i}}, Peyrusse,
  Poy{\'{e}}, Raffestin, Ribolzi, Roth, Schaumann, Serres, Tikhonchuk, Vacar,
  and Woolsey]{Santos-NJP2015}
J~J Santos, M.~Bailly-Grandvaux, L.~Giuffrida, P.~Forestier-Colleoni,
  S.~Fujioka, Z.~Zhang, P.~Korneev, R.~Bouillaud, S.~Dorard, D.~Batani,
  M.~Chevrot, J~E Cross, R.~Crowston, J.-L. J-L Dubois, J.~Gazave, G.~Gregori,
  E~D'Humi{\`{e}}res, S.~Hulin, K.~Ishihara, S.~Kojima, E.~Loyez, J-R J.-R.
  Marqu{\`{e}}s, A.~Morace, P.~Nicola{\"{i}}, O.~Peyrusse, A.~Poy{\'{e}},
  D.~Raffestin, J.~Ribolzi, M.~Roth, G.~Schaumann, F.~Serres, V~T Tikhonchuk,
  P.~Vacar, and N.~Woolsey.
\newblock {Laser-driven platform for generation and characterization of strong
  quasi-static magnetic fields}.
\newblock \emph{New Journal of Physics}, 17\penalty0 (8):\penalty0 083051, aug
  2015.
\newblock ISSN 1367-2630.
\newblock \doi{10.1088/1367-2630/17/8/083051}.
\newblock URL
  \url{http://stacks.iop.org/1367-2630/17/i=8/a=083051?key=crossref.3c00c1af6faa070f173516ea5a628621}.

\bibitem[Sentoku and Kemp(2008)]{Sentoku-jcp08}
Y.~Sentoku and A.J.~J. Kemp.
\newblock {Numerical methods for particle simulations at extreme densities and
  temperatures: Weighted particles, relativistic collisions and reduced
  currents}.
\newblock \emph{Journal of Computational Physics}, 227\penalty0 (14):\penalty0
  6846--6861, jul 2008.
\newblock ISSN 00219991.
\newblock \doi{10.1016/j.jcp.2008.03.043}.
\newblock URL \url{http://dx.doi.org/10.1016/j.jcp.2008.03.043
  http://linkinghub.elsevier.com/retrieve/pii/S0021999108001988}.

\bibitem[Pisarczyk et~al.(2018)Pisarczyk, Gus'kov, Zaras-Szyd{\l}owska, Dudzak,
  Renner, Chodukowski, Dostal, Rusiniak, Burian, Borisenko, Rosinski, Krupka,
  Parys, Klir, Cikhardt, Rezac, Krasa, Rhee, Kubes, Singh, Borodziuk, Krus,
  Juha, Jungwirth, Hrebicek, Medrik, Golasowski, Pfeifer, Skala, Pisarczyk, and
  Korneev]{Pisarczyk-scirep2018}
T.~Pisarczyk, S.~Yu Gus'kov, A.~Zaras-Szyd{\l}owska, R.~Dudzak, O.~Renner,
  T.~Chodukowski, J.~Dostal, Z.~Rusiniak, T.~Burian, N.~Borisenko, M.~Rosinski,
  M.~Krupka, P.~Parys, D.~Klir, J.~Cikhardt, K.~Rezac, J.~Krasa, Y.-J. Rhee,
  P.~Kubes, S.~Singh, S.~Borodziuk, M.~Krus, L.~Juha, K.~Jungwirth,
  J.~Hrebicek, T.~Medrik, J.~Golasowski, M.~Pfeifer, J.~Skala, P.~Pisarczyk,
  and Ph. Korneev.
\newblock {Magnetized plasma implosion in a snail target driven by a
  moderate-intensity laser pulse}.
\newblock \emph{Scientific Reports}, 8\penalty0 (1):\penalty0 17895, dec 2018.
\newblock ISSN 2045-2322.
\newblock \doi{10.1038/s41598-018-36176-8}.
\newblock URL \url{http://www.nature.com/articles/s41598-018-36176-8}.

\bibitem[Bulanov et~al.(2013)Bulanov, {Zh. Esirkepov}, Kando, Koga, Hosokai,
  Zhidkov, and Kodama]{Bulanov-pop2013}
Sergei~V. Bulanov, Timur {Zh. Esirkepov}, Masaki Kando, James~K. Koga, Tomonao
  Hosokai, Alexei~G. Zhidkov, and Ryosuke Kodama.
\newblock {Nonlinear plasma wave in magnetized plasmas}.
\newblock \emph{Physics of Plasmas}, 20\penalty0 (8):\penalty0 083113, aug
  2013.
\newblock ISSN 1070-664X.
\newblock \doi{10.1063/1.4817949}.
\newblock URL \url{http://aip.scitation.org/doi/10.1063/1.4817949}.

\bibitem[Rassou et~al.(2015)Rassou, Bourdier, and Drouin]{rassou-pop15}
S~Rassou, A~Bourdier, and M~Drouin.
\newblock {Influence of a strong longitudinal magnetic field on laser wakefield
  acceleration}.
\newblock \emph{Physics of Plasmas}, 22\penalty0 (7):\penalty0 073104, jul
  2015.
\newblock ISSN 1070-664X.
\newblock \doi{10.1063/1.4923464}.
\newblock URL \url{http://dx.doi.org/10.1063/1.4923464
  http://aip.scitation.org/doi/10.1063/1.4923464
  http://scitation.aip.org/content/aip/journal/pop/22/7/10.1063/1.4923464}.

\bibitem[Hur et~al.(2008)Hur, Gupta, and Suk]{Hur-phla2008}
Min~Sup Hur, Devki~Nandan Gupta, and Hyyong Suk.
\newblock {Enhanced electron trapping by a static longitudinal magnetic field
  in laser wakefield acceleration}.
\newblock \emph{Physics Letters A}, 372\penalty0 (15):\penalty0 2684--2687, apr
  2008.
\newblock ISSN 03759601.
\newblock \doi{10.1016/j.physleta.2007.12.045}.
\newblock URL
  \url{https://linkinghub.elsevier.com/retrieve/pii/S0375960107017872}.

\bibitem[Drouin et~al.(2012)Drouin, Bourdier, Harry, and
  Rassou]{Drouin-jmp2012}
Mathieu Drouin, Alain Bourdier, Quentin Harry, and S{\'{e}}bastien Rassou.
\newblock {Influence of a Static Magnetic Field on Beam Emittance in Laser
  Wakefield Acceleration}.
\newblock \emph{Journal of Modern Physics}, 03\penalty0 (12):\penalty0
  1991--1997, 2012.
\newblock ISSN 2153-1196.
\newblock \doi{10.4236/jmp.2012.312249}.
\newblock URL
  \url{http://www.scirp.org/journal/PaperDownload.aspx?DOI=10.4236/jmp.2012.312249
  http://www.scirp.org/journal/doi.aspx?DOI=10.4236/jmp.2012.312249}.

\bibitem[Jha et~al.(2012)Jha, Saroch, Mishra, and Upadhyay]{Jha-prstab2012}
Pallavi Jha, Akanksha Saroch, Rohit~Kumar Mishra, and Ajay~Kumar Upadhyay.
\newblock {Laser wakefield acceleration in magnetized plasma}.
\newblock \emph{Physical Review Special Topics - Accelerators and Beams},
  15\penalty0 (8):\penalty0 081301, aug 2012.
\newblock ISSN 1098-4402.
\newblock \doi{10.1103/PhysRevSTAB.15.081301}.
\newblock URL \url{http://link.aps.org/doi/10.1103/PhysRevSTAB.15.081301
  https://link.aps.org/doi/10.1103/PhysRevSTAB.15.081301
  https://journals.aps.org/prab/pdf/10.1103/PhysRevSTAB.15.081301}.

\bibitem[Kant et~al.(2016)Kant, Rajput, Giri, and Singh]{Kant-hedp2016}
Niti Kant, Jyoti Rajput, Pankaj Giri, and Arvinder Singh.
\newblock {Effect of axial magnetic field on axicon laser-induced electron
  acceleration}.
\newblock \emph{High Energy Density Physics}, 18:\penalty0 20--25, mar 2016.
\newblock ISSN 15741818.
\newblock \doi{10.1016/j.hedp.2015.12.002}.
\newblock URL \url{http://dx.doi.org/10.1016/j.hedp.2015.12.002
  https://linkinghub.elsevier.com/retrieve/pii/S1574181815001056}.

\bibitem[Kuri et~al.(2018)Kuri, Das, and Patel]{Kuri-lpb2018}
Deep~Kumar Kuri, Nilakshi Das, and Kartik Patel.
\newblock {Collimated proton beams from magnetized near-critical plasmas}.
\newblock \emph{Laser and Particle Beams}, 36\penalty0 (3):\penalty0 276--285,
  sep 2018.
\newblock ISSN 0263-0346.
\newblock \doi{10.1017/S0263034618000307}.
\newblock URL
  \url{https://www.cambridge.org/core/product/identifier/S0263034618000307/type/journal{\_}article}.

\bibitem[Kuri et~al.(2017)Kuri, Das, and Patel]{Kuri-pop2017}
Deep~Kumar Kuri, Nilakshi Das, and Kartik Patel.
\newblock {Proton acceleration from magnetized overdense plasmas}.
\newblock \emph{Physics of Plasmas}, 24\penalty0 (1):\penalty0 013112, jan
  2017.
\newblock ISSN 1070-664X.
\newblock \doi{10.1063/1.4974171}.
\newblock URL \url{http://dx.doi.org/10.1063/1.4974171
  http://aip.scitation.org/doi/10.1063/1.4974171}.

\end{thebibliography}
\bibliographystyle{plainnat}

\end{document}